\documentclass[floats,prd,onecolumn,showpacs]{revtex4-1}
\usepackage{amssymb}
\usepackage{graphics}
\usepackage{graphicx}
\usepackage{amsmath}
\usepackage{amsfonts}
\usepackage{bm}% bold math

\begin{document}

\newcommand*{\be}{\begin{equation}}
\newcommand*{\ee}{\end{equation}}

\title{ Conceptual Problems in Cosmology }

\author{F. J. Amaral Vieira}

\affiliation{\mbox{Departamento de F\'isica, Universidade Estadual Vale 
do Acara\'u (UVA)} \\
\mbox{Avenida Dr. Guarani 317, CEP 62040-730, Sobral, Cear\'a, Brazil} \\
e-mail: amaralvieira@wirelink.com.br }

\date{\today}

\begin{abstract}
In this essay a critical review of present conceptual problems in current 
cosmology is provided from a more philosophical point of view. In essence,
a digression on how could philosophy help cosmologists in what is strictly 
their fundamental endeavor is presented. We start by recalling some examples 
of enduring confrontations among philosophers and physicists on what could 
be contributed by the formers to the day-time striving of the second ones.
Then, a short review of the standard model Friedmann-Lema\^itre-Robertson-Walter
(FLRW) of cosmology is given, since Einstein´s days throughout the Hubble discover 
of the expansion of the universe, the Gamow, Alpher and Herman primordial nucleo-synthesis
calculations and prediction of the cosmic microwave background (CMB) radiation; as 
detected by Penzias and Wilson, to Guth-Linde inflationary scenarios and the 
controversial multiverse and landscape ideas as prospective way outs to most cosmic 
conceptual conundrums. It seems apparent that cosmology is living a golden age with 
the advent of observations of high precision. Nonetheless, a critical revisiting of 
the direction in which it should go on appears also needed, for misconcepts like 
``quantum backgrounds for cosmological classical settings'' and ``quantum gravity 
unification'' have not been properly constructed up-to-date. Thus, knowledge-building 
in cosmology, more than in any other field, should begin with visions of the reality, 
then taking technical form whenever concepts and relations inbetween are translated into 
a mathematical structure. It is mandatory, therefore, that the meaning of such concepts 
be the same for all cosmologists, and that any relationship among all them be tested 
both logically as well as mathematically. In other words, the notorius feature of 
improbability of our universe, as is well-known, assures to cosmologists a priviledged 
degree of freedom for formulating interpretations and theories. However, at the same 
time, it demands for their formulations and conclusions to be considered in the light 
of data taken from astrophysical observations.

\vskip 8.0 truecm

 \vskip 0.5 truecm 
*Revised and extended version of the original article published in the Special 
Volume \textit{A Life of Achievements} - A Festschrift in honor of Prof. Alberto 
Santoro in occasion of his 70$^{\textrm{th}}$ birthday
\end{abstract}

%\begin{flushright}
%\begin{minipage}{8cm}
%Science is not just a collection of facts, it is also understanding; if the 
%understanding does not eventually follow new facts, perhaps there is something 
%wrong with the facts.
%\hfill Michael S. Turner
%\end{minipage}
%\end{flushright}

%\vskip 10.0 truecm

%%%\pacs 

%98.62.En=Electric and Magnetic Fields
%98.80.Es=Observable Cosmology,
%26.60.Gj=...,
%97.10.Cv=...,
%97.60.Jd=....
%\setlength{\maketitle}{-10ex}
 
 \maketitle

\section{Introduction}

The fact that cosmology was able to describe the evolution of the universe starting from a 
time $10^{-43}$~seconds after the creation, {\slshape  i.e.} the Big Bang, to the present time, 
answering some of the most challenging questions proposed by the human mind, has raised the question 
of finding out what contribution philosophers could offer, in our time, to this fascinating field of 
research.

In an approach to the issue, McMullin~\cite{1} quoted Weinberg's answer to this question which would 
become emblematic of the position of a large number of physicists: ``When physicists make discoveries 
in areas that have been the object of philosophy, they not only confirm or refute the speculations of 
philosophers but show that philosophers  were out of their jurisdiction when speculating such problems''

Weinberg, in the chapter ``Against Philosophy'' of his celebrated book {\slshape  Dreams of a Final 
Theory}~\cite{2}, bases his disenchantment with philosophy on the conviction assumed during the days 
of his undergraduate studies ``that the insights of the philosophers he studied 'seemed murky and 
inconsequential' compared with the dazzling successes of physics and mathematics,'' adding: ``From 
time to time, since then, I have tried to read current work on the philosophy of science. Some of 
it I found to be written in a jargon so impenetrable that I can only think that it aimed at impressing 
those who confound obscurity with profundity.'' His conviction is that ``(...) we should not expect it 
(philosophy) to provide today's scientists with any useful guidance about how to go about their work or 
about what they are likely to find.''

Hawking, on his turn~\cite{3}, points out that philosophy has been depleted in as much as ``in the 
nineteenth and twentieth centuries, when science became too technical and mathematical for the philosophers, 
or anyone else except a few specialists.'' To this realistic view of the facts, he opposes an idealistic 
reasoning: ``If we do discover a complete theory, it should in time be understandable in broad principle 
by everyone, not just by a few scientists. Then we shall all, philosophers, scientists, and just ordinary 
people, be able to take part in the discussion of the question on why it is that we and the universe exist.'' 
And through the confrontation between the view of ``how, that is typical of science, and the view of why, 
proper of philosophy'', Hawking comes to a synthesis that may be interpreted both as metaphysical and as a 
product of the most audacious reductionism: ``If we find the answer to that, it would be the ultimate 
triumph of human reason -- for then we should know the mind of God.''

Meanwhile, that is not exactly the question, for since the first decades of the past century the work of the 
cosmologist, starting with Einstein, has been that of systematically transforming their views of reality in 
idealizations called models and of describing such models by means of a logical discourse, a procedure that, 
in essence, corresponds to the method of philosophy. The second and irrefutable step is to base such 
discourse on the laws of physics and to demonstrate it through a mathematical language capable of 
endowing it with the precision and sustainability that characterizes the physicist's work method.

However, how could philosophy help cosmologists in what is strict\-ly their business? The most evident and 
delicate problems of cosmology nowadays are not found solely in the elegance and fineness of its mathematical 
formalisms nor in the technical content of the constructs, but rather in the difficulty to handle concepts.
 And contemporary philosophy, as pointed out by Ara\'{u}jo~\cite{4}, does not have the task of creating 
foundations for the sciences, since this would be a return to the medieval philosophy, while its 
characteristic today is its critical attitude, strict reasoning, ability to handle concepts and integrate 
them, thus having become an articulator of different fields of knowledge.

\section{The Standard Model of Cosmology}

\subsection{The beginning}

Modern cosmology was born in the context of Einstein's General Relativity, in 1915, built around the 
revolutionary idea that gravitation stems directly from the space-time curvature (geometry) and the 
linear momentum of the interacting bodies. In this context, the geometry is described by the metric, 
{\slshape  i.e.} the gravitational field itself. In it, space-time geometry, {\slshape  i.e.}, the 
spatial configuration of the universe as a whole, is described through the so-called gravitational 
field equations, whose form reads

\begin{equation}
R_{\mu\nu} - \frac{1}{2} R g_{\mu\nu} = \frac{8\pi G}{c^4} T_{\mu\nu} \; ,
\end{equation}

\noindent
where $R_{\mu\nu}$ is Ricci tensor, $R$ is the scalar curvature of space-time, $g_{\mu\nu}$ is the metric 
which determines distances in that geometry, $G$ is the gravitational constant, $c$ the speed of light in 
a vacuum and $T_{\mu\nu}$ is the energy momentum tensor of the matter-energy distribution.

Einstein noticed that the solution to those equations led to a dynamic universe which is curved by matter 
and energy, and that could collapse on itself under the effect of gravity. To avert this, for reasons that 
we will discuss later, he amended the field equation in 1917, so as to introduce a new term on the left 
side, which he called the ``cosmological term'', the product of a parameter of dimension ($1/L^2$) times 
the metric. It is designated by the Greek letter $\Lambda$ (Lambda). It would represent an intrinsic property of 
space. Then, the equation were written as:

\begin{equation}R_{\mu\nu} - \frac{1}{2} R g_{\mu\nu} + \Lambda g_{\mu\nu} = 
\frac{8\pi G}{c^4} T_{\mu\nu} \; .
\end{equation}

\noindent Einstein admitted that a positive $\Lambda$ would offset gravity, providing a solution for a 
static universe, {\slshape  i.e.}, one that does not contract nor expand. The result, in summary, is a 
model of universe having a spatially positive curvature, being finite, static, and four-dimensional, 
since it houses space-time as described by the special relativity theory. Nonetheless, Alexander Friedmann, 
until then an obscure Russian physics teacher, between the years 1922 and 1924, derived new solutions 
starting from Einstein's equations which describe three dynamic models of universes simply connected, 
with evolution and destiny related to the specific geometry of each one, namely: (i) a universe with 
Euclidean geometry destined to expand  forever; (ii) a universe with spherical geometry, destined to 
contract under gravitational collapse and (iii) a universe with hyperbolic geometry that would expand 
in an accelerated way. Its original form reads

\begin{equation}
\frac{2R\ddot{R} + \dot{R}^2 + kc^2}{R^2} - \Lambda = 8\pi G \; p
\end{equation}

\noindent
in which $R$ represents the size of the universe, $p$ its pressure, $c$ the speed of light, $G$ the 
gravitational constant, $\Lambda$ the cosmological constant and $k$ the constant of curvature. Thus, 
$k > 0$ defines a curved and closed universe, $k < 0$ defines a hyperbolic universe and $k=0$ defines 
a spatially flat universe, precisely the one we inhabit, according to the most recent data from the 
WMAP satellite.

In 1924, the Belgian astronomer George Lema\^{\i}tre, working independently, arrived at similar 
results. A propos, he had already warned that gravity, as established in the general relativity 
theory, would create a large concentration of matter in the concavities of space-time caused by 
the mass of large astronomical bodies, which contradicted the scenario of visible galaxies uniformly 
spread around. The evolution of Friedmann models follows local physics and the principle postulated 
by Milne, in 1930, according to which the universe is statistically homogeneous, {\slshape  viz.}, 
the same for different observers, placed at different observation points. For the majority of 
authors working in the field, this is a generalization of Copernicus' principle, according to 
which there are no privileged positions in the universe. However, it is necessary to recall that what 
we call Copernicus principle was referred to our solar system, while the homogeneity of the universe, 
as considered today, appears to occur at a very large cosmic scale, {\slshape  i.e.}, beyond 30~Mpc. 
Even after recognizing the accuracy of Friedmann calculations, which demanded time and debates, 
Einstein only admitted that he had committed an error (his great blunder) when he introduced the 
cosmological constant, after Hubble findings in 1929, when he  presented  at the Solvay Congress 
the measurements of the spectrum of the galaxies (distant nebulae), showing that they were moving 
away from us at a speed proportional to distance. Such relation, known afterwards as Hubble's Law, 
is expressed by the equation

\begin{equation}
V = H_0 D \; ,
\end{equation}

\noindent in which $H_0$ is the constant of proportionality between the velocity $V$ and the distance $D$, known 
as the Hubble constant today. The linearity of that law showed that the universe was expanding~\cite{5} 
and made it implicit that going over its history back to the past we would arrive at a point of 
density and curvature tending to the infinite~\cite{6}. This idea of a singular beginning was 
implicit in all of the Friedmann models and had been assumed by Lema\^{\i}tre, who was known to 
express in a letter to the editor of Nature magazine, in 1934, that the universe ``began in a 
state of maximum density on a day without yesterday''.  Hubble's Law, which defines the rate of 
expansion of the universe, can be cast in the simplest form (which is the first Friedmann equation
for $\Lambda = 0$):

\begin{equation} 
H^2 = \frac{8\pi G}{3} \rho - \frac{k}{R^2} \; ,
\end{equation}

\noindent in which the Hubble parameter $H \equiv \left(\frac{\dot{R}}{R}\right)$ is a function of 
time; $\rho$ the density of matter in the universe, $R$ the factor of scale (size) of the universe 
and $k$ defines the curvature of space.

\subsection{Einstein's Great Blunder}

Einstein's mistake in not accepting the dynamic universe revealed by his own equations, changing 
them to make it static, has been the subject of hundreds of articles, essays and debates. Stephen 
Hawking said that such mistake would be understandable if made by Newton who lived two centuries 
before, but thought that Einstein should have been able to see further. However, a judgment is not 
easy. Lee Smolin~\cite{7} admits that even, in those days, Einstein hardly could have imagined a 
universe that were neither eternal nor immutable, simply because that was the view of all scientists 
since Aristotle. He adds, however, that if Einstein were really a genius, he would believe more in 
his own theory than in the prejudice and would have predicted the expansion of the universe. 
Nonetheless, Smolin admits that in his decision Einstein was influenced by the thought that 
predicting an expanding universe may be based on a defect of his theory plus the fact that, 
at the time, there was no evidence in favor of a dynamic universe evolving with time. The 
last argument makes sense. By the time when Einstein introduced General Relativity there was 
the idea that the universe was limited to the Milky Way, (Kapteyn's universe, 1910) where the 
stars seem to move slowly, a view then of serene stability.

In such conditions, it would be inevitable to admit that gravitational attraction would lead 
galaxies to come increasingly closer to one another what would inevitably result in a contraction 
of the universe sometime in the future. It is evident that Einstein took up this scenario as an 
objective fact. Hence, to admit that the expansion foreseen in his equations could be based in 
some drawback or defect of the theory may have been just a small step.

It is undeniable that Einstein was influenced by the ideas of the Austrian physicist and philosopher 
Ernst Mach, who many authors retain might have been the inspiration for the relation between mass 
density and the positively curved geometry established by him, as well as the creation of the 
cosmological constant itself. Physicist Norbert Straumann~\cite{8} agrees that Einstein believed 
that the universe was globally closed because by this time he believed that such a view was the 
only way to satisfy Mach's principle on inertia. He recalls, however, that in a letter to physicist 
F.~Pirani, in 1954, Einstein argued that the positive curvature of space would have been established 
by his own results, which would have prevailed even if the cosmological constant had not been 
introduced. And he added that it would be necessary only to enable a distribution of nearly static 
matter as required by the speed of stars. Taking into account that Mach was a radical positivist, we 
should admit that his influence must have been a guidance for Einstein to take as reference the data 
of the observation relative to the speed of stars and not to influence him toward a vision of a pure 
philosophical nature. But there were many circumstances conspiring in favor of the introduction of 
the cosmological constant, among them the fact that Lambda $\Lambda$ gave to the equations symmetry and 
elegance, something of which Einstein was a devotee.

For Mario Novello~\cite{9}, Einstein amended the field equations to make the gravitational theory 
to fit his mental scenario of the universe, a decision for which there would be no explanation in 
terms of physics. The cosmological constant would be a feature typical of the totality, a 
characteristic of the universe's global structure. Novello adds that this daring step by Einstein, 
without support from any other part of science, must be accepted as the first strategy to free 
cosmology from physics, {\slshape  i.e.}, to allow scientists to be prepared to accept the 
properties that do not have a counterpart in our surroundings.

As pointed out by Weinberg, in inspired and very succinct words, ``Einstein's mistake was that he 
thought it was a mistake.''~\cite{10} To me, it was clearly a conceptual fault as verified 90~years 
later with the discovery of the accelerated expansion of the universe.

\subsection{A long intermezzo}

In the late 1940's, when the idea of a static universe -- despite all that had been established on 
the contrary -- still dominated the preferences of the scientific community, George Gamow entered 
the stage. Although he had been a disciple of Friedmann, he was not directly interested in the 
geometry of the universe, but in the possibility of explaining the creation of the atoms in the 
initial instants of the universe.

The first step was to describe nucleo-synthesis, which required finding a solution to the problem 
of the abundance of some atoms as compared to others, and more precisely the synthesis of the nuclei 
of light elements. Gamow and collaborators Ralph Apher and R.C. Herman demonstrated not the synthesis 
of all the elements, but the synthesis of  ${}^2H_1$, ${}^3He$, ${}^4He$ and ${}^7Li$, in a very 
brief and unique moment, at a time of $10^{-4}$~seconds of the existence of the universe.  At the 
same time, Gamow's team  added to the scenario of creation described by Lema\^{\i}tre the 
inconceivably high temperature that would have made possible all the nuclear reactions necessary 
for creation, for the beginning of everything.~\cite{11} The moment of creation is then understood 
as the cosmological singularity at which the universe's density, temperature and curvature were 
infinite, and at which -- in Barrow's word -- ``everything collided with everything.''~\cite{12} 
That turbulent beginning became known as the Big Bang. With Gamow's contribution emerges the standard 
model of the hot Big Bang or simply the standard model of cosmology (SMC) which is described by the 
metric of Friedmann-Lema\^{\i}tre-Robertson-Walter (FLRW):

\begin{equation}
\mbox{d}r^2 = \mbox{d}t^2 - R^2 (t) \left[ \frac{\mbox{d}r^2}{1 - kr^2}  + r^2 
(\mbox{d}\theta^2 + \sin^2 \theta \mbox{d}\phi^2) \right] \; ,
\end{equation}

\noindent
where $k$, as indicated above defines the type of curvature and $R(t)$ is the factor of scale of the 
expansion~\cite{13}. The dynamics of the expanding universe, and consequently $R(t_{\mbox{rec}})$, 
obeys Einstein's equations through which the expansion rate is related to the content of matter, or 
more precisely to the density of energy $\rho$ and to the pressure $p$. Gamow and his associates 
foresaw the existence of a thermal radiation created in the heat of the Big Bang and that had become 
visible in the recombination phase ($t_{\mbox{rec}}$) when electrons broke away from photons to form 
the first electrically neutral atoms. Free to take the flight of light, the photons would have 
dislocated since that time to the present, in a spatial distribution typical of a black body. 
Nowadays, the radiation predicted by Gamow should be isotropic and fill in the universe in a 
homogeneous manner. Some 20~years later, it would be discovered, at the microwave band, by the 
engineers (today said radio-astronomers) Arno Penzias e Robert Wilson. The Cosmic Microwave 
Background Radiation (CMBR), as it became known, uniformly spreads to all directions in space, 
with the same intensity, showing that it was isotropic and homogeneous, had the spectrum of a 
black body and a temperature of 2.73K (Gamow had predicted a temperature of 5.0K). We can count 
on by now three strong evidences of the Standard Model of Cosmology: (i) the redshift of the light 
coming from receding galaxies which indicated the expansion; (ii) the explanation of the abundance 
of light elements and their synthesis during the initial moments of the universe and (iii) the CMBR 
discovery. And all that was sufficient to bury the stationary model of the universe and the 
cosmological constant.

\begin{figure}[htb]
\begin{center}
\includegraphics[width=0.8\textwidth]{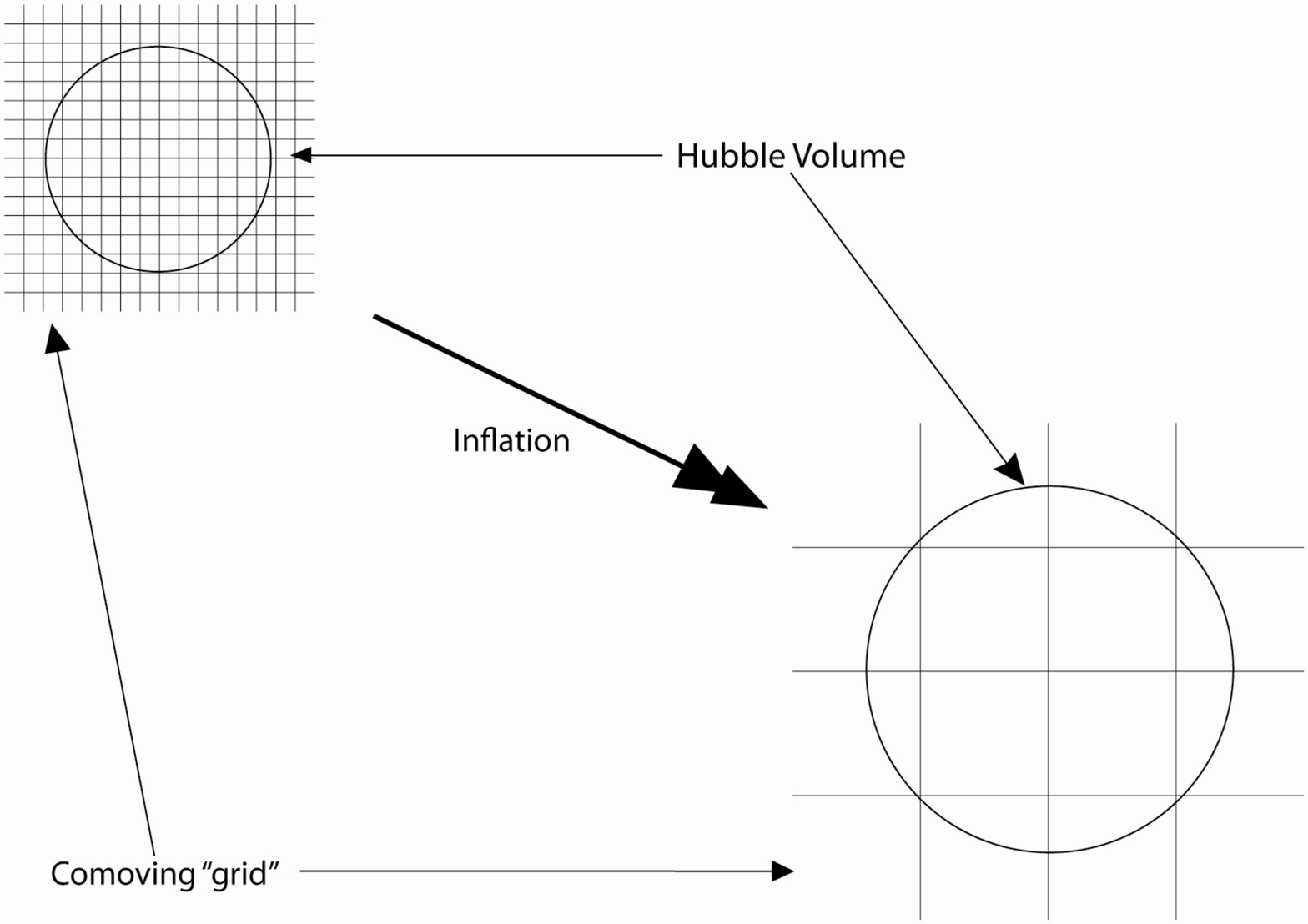}
\end{center}
\caption{The Universe before and after inflation. The gridlines represent the commoving 
coordinate system, and the circle the Hubble volume 1/H, which has shrunk with respect to the 
commoving coordinates after inflation, (From A.H. Jaffe - Cosmology 2010: Lectures Notes - 
Imperial College). (By courtesy of the author).}
\label{fig1}
\end{figure}

\subsection{The Inflationary Expansion}

So much success masked questions for which the Big Bang theory did not have answers as 
we shall see next:

(i) Why is matter distributed homogeneously in large scale just as postulated by SCM? Subsequently, 
the maps of the background radiation obtained from the data collected by satellites COBE and WMAP 
showed that the temperature of the regions symmetrically separated by large distances were homogeneous 
in more than one part in $10^5$ despite not being in causal contact. This difficulty has become known 
as the horizon problem. As pointed out by Brandenberger~\cite{14} it was the fact that regions symmetrically 
separated by large distances from the commoving region $\ell p(t_{\mbox{rec}})$, over  which the CMBR 
is observed to be homogeneous to better than one part in $10^4$, is much larger than the commoving 
forward light cone $\ell_f (t_{\mbox{rec}})$  at the time ($t_{\mbox{rec}}$), which is the maximal 
distance over which microphysical forces could have caused the homogeneity.

(ii) Why is it spatially flat? This peculiarity stems from the coincidence of the value of the total 
density today ($\rho_0$) with the value of the critical density ($\rho_c$), which defines the tenuous 
limit between a negatively curved geometry (open universe) and a positively curved geometry (closed 
universe). The relationship between $\rho_0$ and $\rho_c$ is given by the density parameter $\Omega$ 
such that $\Omega =1$ defines a flat universe, $\Omega > 1$ defines a universe positively curved and 
$\Omega < 1$ defines a universe negatively curved. These relations may be expressed in the form below:

\begin{equation}
\Omega \equiv \frac{\rho}{\rho_{\mbox{crit}}} = \left(\frac{8\pi G}{3 H^2} \right) 
\rho \; ,
\end{equation}

\noindent
where $\rho$ is the density of the universe today and  $\rho_{\mbox{crit}}$ is the critical density. 
Two consequences emerge from what we have just seen: (a) $\Omega =1$ reflects a condition very 
unstable, since a minimal change in their value could change the geometry and the fate of the 
universe; and (b) its value should have been finely tuned in the initial moments of the universe 
and, since then, it has been spatially flat. The theory of the Big Bang had no way to explain one 
thing nor the other.

(iii) Whence could have come the disturbances of density that allowed the formation of agglomerates 
of galaxies? The apparently absolute homogeneity of CMBR, as detected by Penzias and Wilson, did not
offer any leads as to the existence of minor non-homogeneities of temperature so as to allow that 
matter were deposited by gravity in warmer regions from which galaxies could have been formed, while 
the colder regions responded for the space amid them, the voids.

In summary, the cosmology of the Big Bang assumed the homogeneity and flatness, as well as the 
immense size of the universe as determined by initial conditions, of which the most representative 
is the fine tuning required by the value of $\Omega$. But it did not develop a physical mechanism 
to explain them. Big Bang cosmologists faced yet another problem: the high energies of the primeval 
universe would have created super massive magnetic particles that had only one pole, for which reason 
they were called magnetic monopoles, which should exist today in great abundance because they are 
very stable. However, neither were they observed nor physicists envisioned a justification for 
their non-existence today within the standard model of the Big Bang.

It was precisely the interest in magnetic monopoles that led Alan Guth to describe, in 1981~\cite{15}, 
a physical mechanism capable of solving the problems left by the theory of the Big Bang, which 
established an indissoluble link between relativistic cosmology  and quantum theory.  The hypothesis 
underpinning Guth's model is that during the infinitesimal interval of time of $10^{-34}$ seconds the 
scale factor $R(t)$ had grown exponentially, sustaining this up to $10^{-32}$ seconds, 
allowing for the size of the universe to rise from $\sim 10^{-26}$~m to something around 
1~meter, this way growing proportionally to the distance scale it reached over all of its 
existence hitherto.  This brief and violent period of inflationary expansion, as named by 
Guth, brought an elegant and convincing explanation to the problems that we have just 
discussed.~\cite{16}

\begin{figure}[htb]
\begin{center}
\includegraphics[width=0.8\textwidth]{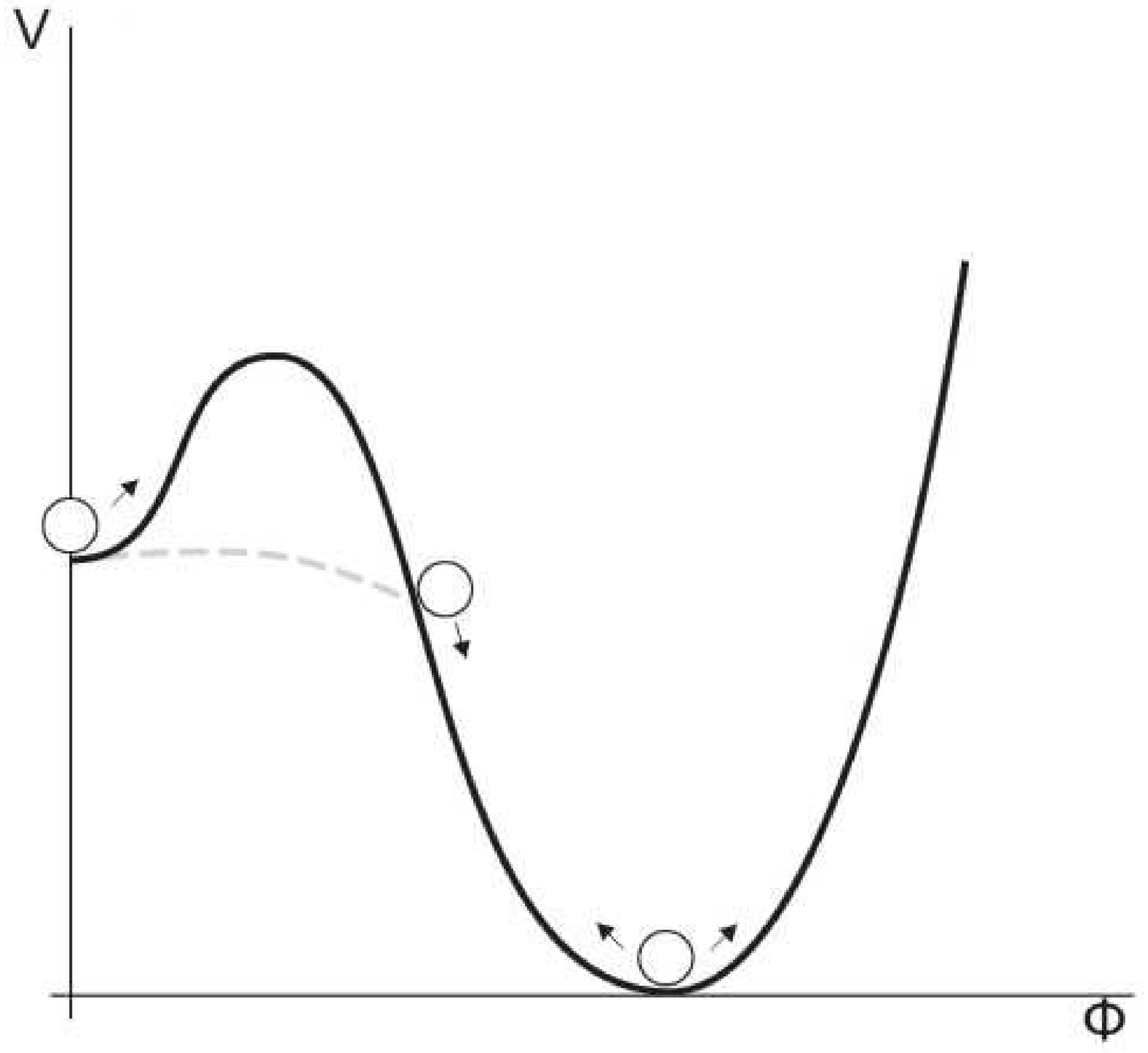}
\end{center}
\caption{Evolution of Scalar Field in Old Inflation. The field  tunnels a barrier of 
energy   and  ``rolls''down toward the minimum where it oscillates.}
\label{fig2}
\end{figure}

As pointed out by Jaffe~\cite{17},  during inflation the Hubble scale remains constant, but 
commoving scales increase much more rapidly, that is, the true horizon scale is now much larger 
than in Hubble length. (See Fig.~\ref{fig1}). Thus, all scales that were compressed within the 
horizon before the inflationary expansion have remained outside the horizon at $t_{\rm rec}$ 
when CMBR became visible. In other words, inflation has placed in causal contact all regions of 
the universe before they were separated by distances beyond the cosmic horizon. As it can easily 
be deduced, during inflation the cone of light for the future increased exponentially, while the 
same did not happen in relation to the past light cone. As we see, inflation cancelled the horizon 
problem, too.

The solution to the problem of flatness seemed more obvious and is a natural consequence of an 
exponential increase of $R(t)$. The effect of gravity, as pointed out by Guth~\cite{18} is reversed 
during the inflation, changing all the equations that describe the evolution of the universe so that, 
instead of moving away from 1, $\Omega$ is brought to 1 at incredible speed, adding that in 100 times 
of duplicating the difference between $\Omega$ and 1 decreases by a factor of $10^{60}$. If we can 
expand a curved surface, continuously and for sufficient time, there will come the time when it will 
appear to be flat in relation to the coordinates in which it is defined. This is precisely what we 
see as the surface of our planet and as we imagine should occur with the three dimensional ``surface'' 
of the universe. On the other hand, all scales that were compressed within the horizon before the 
inflationary expansion have remained outside the horizon by the trec when CMBR became visible. In 
other words, inflation has placed in causal contact all regions of the universe before they were 
separated by distances beyond the event horizon. Villela, Wuensche and Leonardi~\cite{19} showed 
that one may reach such a conclusion, through a mathematical procedure, by comparing the distance 
that light can travel at the beginning of inflation $t_{\mbox{inf}}$, during an interval $\Delta t$, 
with the distance that it could have traveled after recombination. As is ease to deduce, during  
inflation the cone of light for the future increased exponentially, while the same did not happen in 
relation to the past light cone. As we see, inflation has removed the horizon problem too. This same 
homogeneizing power of inflation would dilute the magnetic monopoles making its present number 
irrelevant.

To build his model of inflationary expansion, Guth admitted that the primeval universe, at the 
time of creation, would have been dominated by a type of energy predicted in high energy physics 
capable of exerting a negative pressure and, therefore, to offset gravity. That energy, different 
from all that cosmologists had encountered, Guth ended up finding it in the quantum vacuum emerging 
as a scalar field (Higgs field) in the form of a density potential $V(\phi)$. This potential, namely 
false vacuum, was defined as a peculiar form of matter, a metastable state, produced by a phase 
transition of super-cooling. It decays through bubbles nucleation until it reaches its minimum value 
and start to oscillate, the point at which the universe is being reheated. The electrical barrier is 
part of the characteristics of the diagram of energy employed. (See Graphic~\ref{fig2}).

For what matters to us, it has a negative pressure $\rho =V(\phi)$, which, being opposed to gravity, 
creates a rapid exponential expansion state while it is sustained. Guth found the technical bases 
and mathematical formalisms for its model in the Grand Unified Theories (GUTs).

\begin{figure}[htb]
\begin{center}
\includegraphics[width=0.8\textwidth]{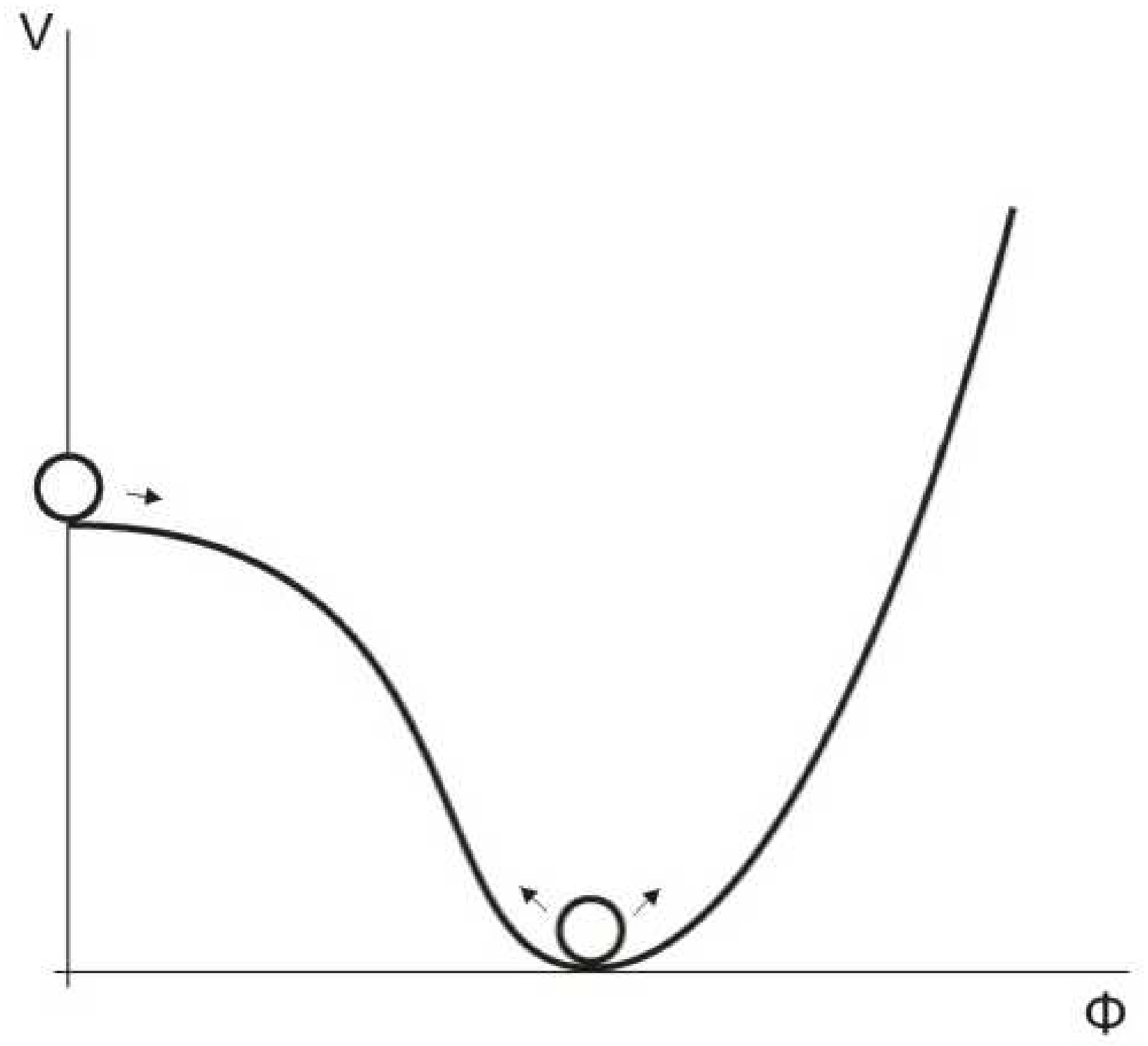}
\end{center}
\caption{Evolution of Scalar Field in New Inflation. The field ``rolls'' down slowly  
toward the minimum where it oscillates.}
\label{fig3}
\end{figure}

Guth had the merit of introducing a phase of exponential expansion in the initial moments of the 
universe, with technical bases and conceptually comprehensible, establishing its technical 
foundations and defining its role in the universe's evolution. His inflationary scenario would be 
of a statuesque beauty were it not for two  difficulties:  (i) the first of a conceptual nature, 
which is summed up in that it does not solve the fine tuning problem; (ii) the second of a technical 
nature, which lies in the fact that inflation did not end in the direction predicted by his model. 
Guth explained that when inflation comes to an end, bubbles of matter are formed and grow, while the 
energy is released. But, rather than distributing itself uniformly through space, that energy 
increasingly concentrated within the walls of the bubbles as they expanded.  Thus, it is only when 
the bubbles collide that energy may spread uniformly throughout the universe in the form of squirts 
of particles in all directions. At that point, however, a complication appeared: as the walls of 
bubbles move apart at the speed of light -- and nothing can move more quickly -- the effects may not 
pass from its own wall while the space continues to expand exponentially. In short, the expansion 
itself avoids collision between the bubbles and this detail prevents the formation of the soup of 
hot particles from which everything would be made. The inflationary model was elegant, simple and, 
in general lines, conceptually correct, but it did not work properly as a result of a technical 
problem whose solution was called "graceful exit." That solution was first found by Andrei 
Linde~\cite{20} of Stanford University and, soon afterward, by Paul Steinhardt and Andreas Albrecht 
of the University of Pennsylvania.~\cite{21} Summarizing, they introduced changes in the equations 
so that the energy of the false vacuum did not decay so rapidly, as it was the case in Guth's model,
 but slowly, which means, in practice, altering the original energy diagram to attain a smooth 
transition of phase. (See Fig.~\ref{fig3}). With such change, when the energy oscillates as it 
reaches true vacuum, it gets converted into the hot and uniform soup of electrons, gluons and 
quarks from which everything was to be made.

\subsection{From Eternal Inflation to Multiverse}

\begin{figure}[htb]
\begin{center}
\includegraphics[width=0.7\textwidth]{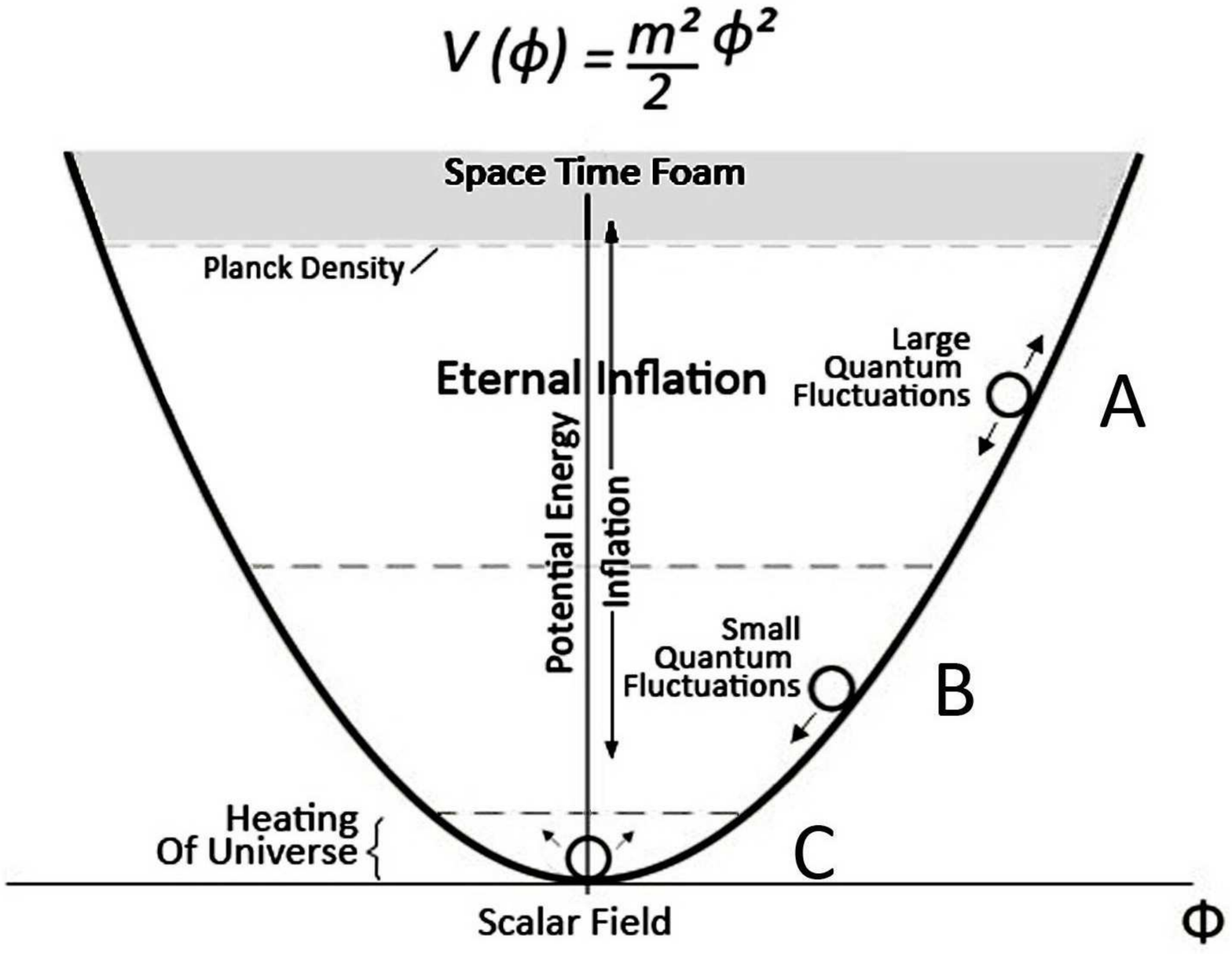}
\end{center}
\caption{Evolution of the scalar field in the theory with $V(\phi) = m^2\phi^2/2$ 
(According to Linde and by his courtesy.)}
\label{fig4}
\end{figure}

In 1983, Linde invented ``the Chaotic Inflation''. The new model, according to him, does not 
require an initial state of thermal equilibrium, super-cooling and tunneling from the false 
vacuum and it appears in theories that can be as simple as that of the harmonic oscillator
\cite{22}.  Its descriptive paradigm invokes a scalar field $\phi$, with mass $m$ and potential 
energy density $V(\phi)=1/2m^2\phi^2$   that rolls from the high value in region $A$ (see Fig. 
\ref{fig4})  toward a minimum of $V(\phi)$ in region $C$, when it oscillates creating pairs of 
elementary particles and drives the universe heating.

\begin{figure}[htb]
\begin{center}
\includegraphics[width=0.8\textwidth]{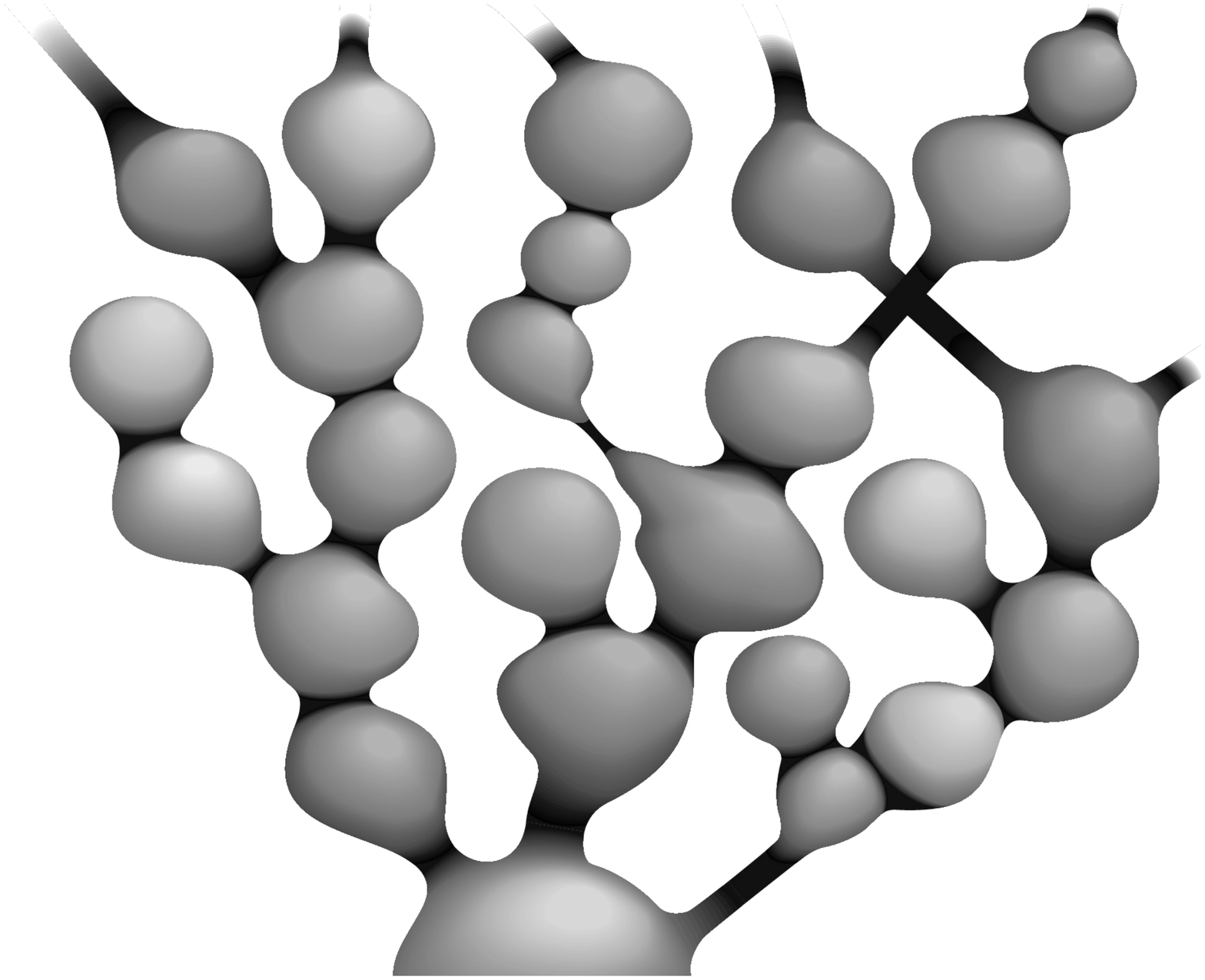}
\end{center}
\caption{The multiverse scenario - A surrealistic tree of inflationary bubbles illustrate 
different domains  with different initial conditions and properties. (According to Andrei Linde 
and by his courtesy.)}
\label{fig5}
\end{figure}

It should be recalled that the large quantum fluctuations that fed inflation occurred not at 
a specific point, but at different parts of the universe, in jumps of a scalar field. Those 
particularities lead to the eternal process of self-reproduction of the universe. As Linde 
states, if the Hubble constant during inflation is sufficiently large, quantum fluctuations 
of the scalar fields may lead not only to the formation of galaxies but also to the division 
of the universe into exponentially large domains with different properties.  The consequence 
is that our universe is no longer everything that exists, but a finite part of a huge and 
eternal universe, that is, the multiverse. And this is the more important issue of Linde's 
model. In Figure~\ref{fig5}, created by Linde, one can see a collection of bubbles interconnected 
by isthmuses. Thus, it might suggest a permanent connection between them. This scenario may lead 
one, notwithstanding, to consider that as far as the baby universe expands there exists a high 
probability that the istmuth through which it is connected to the mother universe, gets so 
stretched till the it gets broken apart, so as is figured out by Guth when describing the creation 
of universes at Lab conditions \cite{15}. That why I posed the following question to Linde whether 
we should not consider the possibility that a baby universe might detach itself from the original 
universe in 
microseconds following its creation. Linde explained that in his approach different ``universes'' 
are simply very large separate parts of the same universe and they are so far from each other that 
inhabitants of one of them do not know anything about the other part of the universe. And he adds, 
one can approximately consider them separate and independent, but in fact they are a part of the 
same manifold and they cannot separate from it. This our interpretation leads, in the meantime, to 
topological and conceptual problems of extremely difficult solution. In this context, it is interesting 
then to notice 
that Garcia Aspeitia and Matos \cite{tonatuih2010} suggest that in order to achieve a consistent 
description of the universe dynamics upon topological considerations it would be required to immmerse
the universe in an (5-dim) extra-dimensional manifold. Hence is that simple to perceive that we 
are dealing with an extremely complex problem. 

Linde added that ``the figures illustrating this concept are very imperfect because they are an 
attempt to show curved space, which is very hard to do without creating some confusion.'' He 
further asserted that ``At the level of quantum cosmology, one can talk about many totally 
separate universes. However, this possibility is much more difficult to study, so most people 
now concentrate on investigating the simpler picture when huge universes (or multiverses) 
consist of many smaller but still huge ``universes''.''

Based on Linde's explanation, we should avoid the idea that our universe is an ``island universe'' 
inside huge archipelagos of connected worlds, since this could lead to the conceptual problem of 
knowing if there is an external space for each particular universe that was not yet studied in 
technical terms.

The biggest problem of the post-Big Bang Theory has been the fine tuning of constants, parameters 
and numbers associated to the initial conditions of the universe precisely because all of them are 
directly related to the evolution of life and, particularly, of intelligent life, which were studied 
in detail by John Barrow~\cite{23}. Physicists and cosmologists broke up into those who preferred to 
admit that they were dealing with a fine tuning that was only in part apparent and those convinced 
that the problem was already solved by classical inflation. Linde was among those who tackled the 
problem head on.~\cite{24}  In his book published in 1990~\cite{25}, he offered an idea of the 
problem with some examples: ``by increasing the mass of the electron by a factor of two and one-half 
would make it impossible for atoms to exist; multiplying   $\alpha_e$ by one and one-half  would 
cause protons and nuclei to become unstable; and more than a ten percent increase in $\alpha_s$ 
would lead to a universe devoided of hydrogen. Adding or subtracting even a single spatial dimension 
would make planetary systems impossible, since in space-time with dimensionality $d > 4$, 
gravitational forces between distant bodies fall off faster than $r^{-2}$, and in space-time with 
$d < 4$, the general theory of relativity tells that such forces are absent altogether''. He 
concludes by recalling that to shelter life such as we know it, ``it is necessary that the 
universe be sufficiently large, isotropic and homogeneous.'' Such a large number of delicately 
tuned figures that are directly related to life cannot be attributed to mere coincidences, 
especially as we have already stressed, as they are unequivocally related to the emergence of 
intelligent life. This conviction leads directly to the anthropic principle introduced by 
Brandon Carter~\cite{26}, which is essentially a way out envisaged for limiting the 
principle of Copernicus. ``En passant´´, let us mention that Carter defended that idea 
at a conference in Krakow, Poland, in 1974, 
as part of the celebration of the 500 years of Copernicus, which is rather ironical. The 
(Copernicus) principle became a matter of mandatory consideration by cosmologists following 
the publication of the book by John Barrow and Frank Tipler~\cite{27} (indeed, not just a 
book, but a {\slshape  vademecum} on the initial conditions of the universe and its future 
from the view of the anthropic principle.)

In the present case it is not a question of rejecting a theory, but simply of knowing if the 
explanation contained in it yields an answer that is adequate to the problem that it intends 
to eliminate.

The solution of the problems of homogeneity and flatness is easily accepted because a set of 
reliable observations shows that the universe went through a phase of exponential expansion. 
But there is a great discrepancy between, on the one hand, the deliberate fine tuning of the 
values of a large number of parameters, as well as their linear relation with the existence 
of life, and on the other hand an extremely complex solution that is not subject to observation. 
One must remember that, in the vicinity of a universe, including ours, there must coexist regions 
of space that are inflating at the speed of light, which eliminates any possibility of an expedition 
to another particular universe governed by physical laws that in all probability are different from 
ours.

And the problem is not only conceptual -- although this is a very important aspect -- but also 
technical, as recognized by Stoeger, Ellis and Kischner~\cite{28}. They argue that ``one way one 
might make a reasonable claim for the existence of a multiverse would be if one could show that 
its existence was a more or less inevitable consequence of well-established physical laws and 
processes. Indeed, this is essentially the claim that is made in the case of chaotic inflation. 
However, the problem is that the proposed underlying physics has not been tested, and indeed may 
be untestable. There is no evidence that the postulated physics is true even in this universe, 
much less in some pre-existing metaspace that might generate a multiverse''.

Jim Hartle and Stephen Hawking~\cite{29} tried to describe a wave function of the universe, considering all 
of its possible histories, which is similar to considering that all universes are possible. And 
the wave function they found is concentrated in our universe, assuming infinitesimal values for 
all other universes. Thus, Linde's model, despite its elegance, reasoning and a certain air of 
completeness, and the Hawking model, despite its daring endeavor, seem to bring into reality 
only one universe, precisely that in which we live in and whose existence cannot be put into 
question.

\subsection{From the Big Bang to the Big Bouncing }

General relativity (GTR) equations converge to a point when the history of the universe 
is followed backwards to the beginning wherein the $space$ + $time$ metric (space-time) 
becomes singular and the universe temperature and density become infinite. As postulated 
by Hawking: ``Time begins at the Big Bang and ends on Black Holes''\cite{hawking-penrose1996}. 
Notwithstanding, he 
recalled very recently in a public lecture at Universeit von Leuven, time is fundamentally 
different from spatial directions, and that it only behaves as any other spatial coordinate 
under extreme conditions, like in the beginning of the universe, near the Planck time. This 
is equivalent to say that in GTR time, itself, is \textit{always} a dynamical coordinate, as 
seen for instance for a stationary observer seing some clock falling into a black hole and 
reaching its horizon, or at the Big Bang, while the pure spatial ones do not exhibit such 
feature. In fact, they are always non-evolving coordinates, as in Newtonian mechanics.

At this point,
the so-called "initial singularity``, the equations of GTR collapse and space-time 
disappears. Such singular origin of the universe is considered by plenty of cosmologists 
a very serious embarrasing and unsuspected drawback, which may justify to reject the 
standard model of cosmology despite the recognized success and observational verification 
of its predictions. This led to look for building alternative models in which the singularity
would be eliminated or somehow avoided, as is the case of the so-called cyclic universe 
models in which each phase of contraction is followed by another one of accelerate expansion,  
being this phase reversal determined by a quantum mechanism, that is as energetic as the 
Big Bang itself, generically named as ``bouncing''. It is stressed that the bouncing is 
one of the plenty of mechanisms figured out to overcome the intial singularity problem
(string theory, M-theory, Smolin's cosmological natural selection -bouncing black holes-), 
varying speed of light, etc.), and all of them are admitted by the most general theory 
unifying gravity and quantum physics.  

Regarding this field of research, there is a fact: part of the community of cosmologists 
working in this branch makes use of a combination of fundamentally different concepts in 
formulating most of the quantum effects-inspired bouncing theories. Indeed, Hawking 
created his model of unbound universe in which the initial singularity is avoided based 
on a quantum theory of gravitation that has not been formaly developed, as himself admits
in his book \cite{Hawking-BHoT}: 

``We don't yet have a complete and consistent theory that combines quantum mechanics and 
gravity -\textit{Note of the author: vis-a-vis general relativity}-. However, we are 
fairly certain of some features that such a unified theory should have. One is that it 
should incorporate Feynman's proposal to formulate quantum theory in terms of a sum over 
histories. In this approach, a particle does not have just a single history, as it would 
in a classical theory. Instead, it is supposed to follow every possible path in
space-time, and with each of these histories there are associated a couple of 
numbers, one represent-ing the size of a wave and the other representing its 
position in the cycle (its phase). The probability that the particle, say, passes
through some particular point is found by adding up the waves associated with every 
possible history that passes through that point. When one actually tries to perform 
these sums, however, one runs into severe technical problems. The only way around 
these is the following peculiar prescription: one must add up the waves for particle
histories that are not in the "real" time that you and I experience but take place 
in what is called imaginary time. Imaginary time may sound like science fiction but 
it is in fact a well-defined mathematical concept. If we take any ordinary (or "real") 
number and multiply it by itself, the result is a positive number. (For example, 2 
times 2 is 4, but so is -2 times -2.) There are, however, special numbers (called 
imaginary numbers) that give negative numbers when multiplied by themselves. (The one 
called ``i'', when multiplied by itself, gives -1, 2i multiplied by itself gives -4, 
and so on.) ... '' 
 
The current status on the search for such a theory can be illustrated, as a matter of 
example, by quoting arguments extracted from the review article by Pinto-Neto 
\cite{pinto-neto2011}: 

``Any mechanism which eliminates the initial singularity yields at most two possible 
scenarios: either there is a quantum creation of a small but finite universe and hence 
a beginning of time, or the universe is eternal, with no beginning of time. This last 
possibility can be divided in two other cathegories: the universe has always expanded, 
with a long accelerated phase before the usual decelerated expansion of the standard 
model from either an asymptotically zero volume flat space-time or from a finite but 
small compact region; the universe had a contracting phase before the usual expanding 
phase, hence performing a bounce. The transition from accelerated to 
decelerated expansion or from contraction to expansion demands non standard or non 
classical physics (modifications of general relativity through nonminimal couplings, 
non-linear curvature in the lagrangian, quantum effects in matter fields and/or 
gravitational fields, etc., and/or non-standard matter (N-of-the-A: i. e., matter 
violating GTR energy conditions) in order to avoid the singularity in between''. 
In virtue of this state-of-the-art, it is timely to recall that independently of 
the specific mechanism that comes to be postulated, any collapse without a perfect 
simmetry should produce a bouncing, as shown by Lifshitz and Khalatnikov \cite{L-K1970} 
in the quasi-isotropic solution of Einstein equations near cosmological singularity
of complex and oscillatory nature, for the case of arbitrary hydrodynamical matter.

It becomes apparent that if one critically analyzes the set of fashions proposed to 
overcome the singularity problem, one realizes immediately that three of them no needs 
fundamental changes to the physical role of the gravitational field if understood as a 
pure space-time curvature phenomenon. Meanwhile, invocation of quantum effects in 
matter and/or gravitational fields implies, in the first case, to superpose to the 
classical gravitational field a number of quantum effects. However, such method clearly 
means to attempt to fuse together physical theories that are fundamentally different 
from the conceptual point of view, as depicted by Eq.(\ref{action-1})  

\begin{equation}
S = \int{d^4\rm{x} \; (R + L^{Matter}_{Quantum-Effects})} \;  .
\label{action-1}
\end{equation}

Whilst for the second proposed procedure, such way out implies to have in 
hands a quantum theory of gravity. Unfortunately, to overrun this conceptual 
dicothomy such a construct till to date has not been unveiled, if indeed there 
exists such a thing, as many workers in the field have questioned. 

From an alternative perspective, and returning to the bouncing black holes 
singularity quoted above, lots of authoritative authors, including the reverencied 
book on \textit{Gravitation} by Misner, Thorne and Wheeler \cite{MTW1973} who have 
suggested that perhaps we may be inhabitating a universe which 
resembles (is) in most respects a classical (Schwarzschild-like) black hole. They 
have also demonstrated that the coordinate system used to describe the dynamics of 
an expanding universe can  also  be used to describe univoquely the gravitational 
collapse of a mass leading to the formation of a black hole, and its forever 
``censured'' singularity, after simply reverting the direction of evolution of the 
time coordinate in the system being used. Hence, it seems that if such a view is 
correct, then the singularity issue gains a new dimension as regarding the transition 
from a contraction to an accelerate expansion phase, since it appears to be implicit 
that everything that exists inside this universe had come out from such singularity.
Such a conclusion by itself conveys once again conceptual problems. 

Summarizing, the recent revision by Novello and P\'erez Bergliaffa in 
``Bouncing Cosmologies'' \cite{novello} shows that the bouncing models took new 
impetus after the discovery of the acceleration of the universe and should have 
a room in the cosmological debate. 

Hence, the expectation proceeds: whether the value of Lambda has changed over the cosmic time 
so as to allow for a phase of exponential expansion, followed by phase of deceleration 
that drove the acceleration phase we are undergoing by now, then, it appears reasonable 
to imagine that the same mechanism of cancellation might produce a deceleration phase 
able to drive the universe to a contraction stage. Within this line of reasoning, the 
reversal of the contraction phase to a phase of exponential expansion is fairly similar 
to the inflationary expansion which constituted by itself the central event of the 
standard model Lambda-CDM. A model exhibiting such characteristics, which properly 
could be denominated the \textit{Lambda-Bouncing Cosmology}, should accommodate the 
inflationary expansion phase, and thus fully describe the mechanism for the reversal 
of the contraction stage, in a time scale able to drive the universe throughout the 
stages that are current proved to show consistency with most of current astrophysical 
observations. 

\subsection{The resurrection of the Cosmological Constant}

In 1998, two research groups working independently - the Supernova Cosmology Projetct, 
led by Saul Permutter~\cite{30} and the High-Z Supernova Search Team, headed by Adam 
Riess \cite{31} announced 
that the expansion of the universe had undergone an acceleration at a recent cosmic time. The method 
that these group employed  consisted in locating supernovae of type IA (SNIA), the light curve of 
which appear to evolve as standard candles, and are situated billions of light years away. From 
their intense brightness astronomers can calculate the distante and the time the SNIA light took 
to arrive to Earth from its location. As soon as SNIA event is detected, astronomers look for the 
measurement of the redshift of the (host) galaxy found at each field of view   around a SNIA. 
Basically, it was observed that the distance increased  with the redshift  much more rapidly than 
was expected in the context of the standard cosmological model. The discovery has been widely 
confirmed by data obtained from the CMBR collected by the Wilkinson Microwave Background Probe 
(WMAP), whose maps show that dark energy represents around 74\% of the energy density 
of the entire universe (dark matter represents 22~\% and the intergalatic gas and dust
 3.6\%.

Nikcholas Suntzet, astronomer of the group of Adam Riess, once said that: ``What we have found is 
that there is a `dark force' that permeates the universe and that has overcome the force of gravity. 
This result is so strange and unexpected, that it perhaps is only believable because two independent 
international groups have found the same effect in their data.'' Soon, it was perceived that the 
dark energy had all the characteristics of the  vacuum energy, and that it was mathematically 
equivalent to it.  This led directly to the so-called Problem of the Cosmological Constant summed 
up in the fact that the attempts to establish the value of the energy of the vacuum -- the energy 
of point zero -- lead to divergent values.

Robert R. Caldwell pointed that if the vacuum energy density really is so enormous as predicted by quantum 
physics it would cause an exponentially rapid expansion of the universe that would rip apart all 
the electrostatic and nuclear bonds that hold atoms and molecules together, and so there would be 
no galaxies, stars or life~\cite{32}.  So that we must admit the existence of a miraculous 
cancellation mechanism that instead of making Lambda exactly zero, only cancels it to 120 
decimal places.  Those requirements, says Caldwell, seem bizarre.  And he explains why: ``Some 
constant that is naturally enormous must be cut down by 120 orders of magnitude, but with 
such precision that today it has just the right value to account for the missing energy''. 
But, this is precisely what was required when the time came for the formation of galaxies, 
stars and planets, such as Earth, capable of harboring life and observers.

The solution to this impasse was proposed by Weinberg in 1987~\cite{33} with the introduction 
of the Anthropic Principle, in its weak version, understood by him as an explanation ``of which 
of the various possible eras or parts of the universe we could inhabit, by calculating which eras 
or parts of the universe we could inhabit''. ``Desesperate situations require desesperate measures, 
and in 1987 Steven Weinberg, one of the most eminent scientists in the world, acted in desesperation,
'' said Leonard Susskind~\cite{34}.  He is right because, at the time, the anthropic principle was a 
forbidden expression for most physicists.

The discovery of the cosmic acceleration not only resuscitated the Cosmological Constant Problem, 
but made it even more challenging since it showed that the value of the energy of the vacuum 
increase over a time that coincides with our presence at this point of the universe.  This is 
also namely the problem on the cosmic coincidence.

 In summary, the high value of the cosmological constant that we will find at inflation was 
drastically reduced to the time it took for the formation of the large structures, and then 
varied again, this time increasing over a time that coincides with the emergence of life and 
consciousness.

Yakob Zeldovich, in 1967~\cite{35}  and in 1968~\cite{36}, was the first to relate, so convincingly, 
the cosmological constant and the energy of the vacuum, so that $\Lambda =8\pi G \rho_{\mbox{vac}}$. 
This idea though well founded was not easily assimilated for reasons that Zeldovich and I. D. Novikov 
would explain in a book of great importance to cosmology published in 1983~\cite{37} in the 
following terms:

\begin{quotation}
``Give the tendency toward objectivity and completeness, this practice would already have 
been a reason enough for the theory with a zero $\Lambda$ to be located  -- at least  in 
small type -- in an appendix. But sensible, calm decisions  are often made only under the 
pressure of extraordinary  circumstances, in a situation of fire and/or flood and 
panic.''
\end{quotation}

That is so much so that Guth, in his inspired book, to which we have referred to early, published 
in 1998, admits that the gravitational effect of the false vacuum during inflation is identical to 
the effect of the cosmological constant. Nonetheless, he recalls the existence of an important 
difference between the two: ``While the cosmological constant is a permanent term in the universal 
equations of gravity, the false vacuum is an ephemeral state that exercised its influence but for 
a rather brief moment at start of history in the past.''

Notwithstanding, time has shown that the cosmological constant is the value we attribute to the 
energy of the vacuum that has varied in time apparently in order to support life. In this context 
it is opportune to quote from a comment by Jayant V. Narlikar and Geoffrey Burbidge~\cite{38}:

\begin{quotation}
``This is ironical, since the one reason for invoking inflation was to avoid fine tuning of 
precisely this nature.  Now it appears that inflation brought its own fine tuning to an even 
greater degree!''
\end{quotation}

The solution to what I will call the Expanded Problem of the Cosmological Constant was the same 
proposed by Weinberg in 1987, the anthropic principle, today widely employed in different 
cosmological models, as Linde's Multiverse and the Susskind's Landscape, created in the context 
of the string theories. This model fits the consideration and test of a variety of hypotheses 
since it is not a real place, but a space of possibilities, a mathematical construction, as 
defined by Susskind. In it all properties and environments and a number of dimensions that 
could reach hundreds and thousands are present. The landscape is undoubtedly a field of tests 
of the anthropic principle.

There are other scenarios and models for dark energy (in addition to the energy of the vacuum) 
within which we could tackle the expanded problem of the cosmological constant, as that of the 
Quintessence and Modified Gravity. (See Frieman, Turner and Huterer~\cite{39}). 
The energy of the vacuum, however, is the most plausible and the simplest to solve the 
problem. But without the anthropic principle it becomes a most complicated puzzle.

\section{Discussion and final remarks}

L. Krauss and M. Turner~\cite{40}, commenting what they called ``a cosmic connundrum'', 
recall that the term cosmological, today called cosmological constant, resuscitated 
thanks to the evidence of the expansion of the universe and directly from the principles 
of quantum physics, the branch of physics that Einstein so famously abhorred. Indeed, 
the mysterious form of energy that the cosmological constant represents emerges in the 
quantum vacuum through the continuous generation of the so-called virtual particles, 
admitted by the uncertainty principle, which come into reality in pairs, but which 
eliminate each other mutually before they are detected. The most current version of 
the standard model of cosmology, nowadays named $\Lambda$-CDM, accommodates all of the 
results gathered from the CMB radiation and those from the observations of SNIa, specillay 
dark energy is interpreted as vacuum energy, that is $\Lambda$. Such priviledged status of 
the standard model $\Lambda$-CDM is emphasized by Mike Turner \cite{MT1997}: ``The case is 
simple'': there is no compelling theoretical argument against a cosmological constant,  
and $\Lambda$-CDM is the only CDM model that is consistent with all present observations. 
$\Lambda$-CDM has two noteworthy features: it can be falsified in the near future (the 
prediction $q_0 \sim -0.5$ is an especially good test), and, if correct, it has important 
implications for fundamental physics''. Notwithstanding, the standard model has yet to 
solve both the fundamental and very serious shortcomings it is facing presently: 
the cosmological constant problem (a $\Lambda$ value today at least $10^{56}$ 
times larger than observations allowed) and the coincidence problem (one among 
several of the so-called \textit{fine tuning} problems).

The first step is to consider that the Anthropic Principle constitutes the unique 
available instrument that we have to provide an explanation of the fine tuning of the 
parameters associated to the initial conditions of the universe, including the value 
of $\Lambda$, and leading to identify their attractors. Nonetheless, over decades the 
Anthropic Principle was treated as something anti-scientific, what restricted its 
usage to the creation of probabilistic stratagemas in the context of the multiverse 
proposal. The anthropic principle, the models of multiverse, the cyclical universe 
and many other ideas have faced this type of restriction in the last decades at the 
same time that the use of increasingly sophisticated mathematical techniques seemed 
to make unnecessary the logical discourse and the systematic conceptual criticism. 
This represents one of the most serious conceptual errors in the approach to the 
problems of cosmology in our days, namely, the mistake of establishing what must or 
musn't be accepted as explanation and even as working hypothesis, which amounts, in 
essence, to choosing which properties can the universe have or not have in the realm 
of a model. In this very respect, just this year, Barrow and Shaw in 
\cite{Barrow2011} pointed 
out that: ``Attempts to explain the coincidence that $\Lambda \sim 1/t^2_U$ ($t_U$ 
is the age of the universe) have relied upon ensembles of possible universes, in 
which all possible values of $\Lambda$ are found. Anthropic selection is combined 
with some prior probability distribution for $\Lambda$ over the ensemble to find 
the most probable value (\textit{N-of-the-A: These authors seem to refer to the
attractors quoted in the precedent discussion}) that allows galaxies to form. Clearly, 
it would be much more attractive to predict $\Lambda$ directly using a testable 
theory without appeal to a multiverse of possibilities ...''. Hence, it is not 
negligible the possibility that the Anthropic Principle may lead to the 
formulation of a theory with such purpose. 

Times seem to be changing. This is what emanates from the book {\slshape Universe or 
Multiverse?} published by Bernard J. Carr in 2007~\cite{41}. It brings together eminent 
cosmologists, physicists, and philosophers who are all doing good philosophy. In the 
remissive index of the book the word ``Anthropic'' appears at least 78 times, many more 
than any other. The delicate, complex and not yet fully assimilated theme of the multiverse 
is discussed in all its versions. And the focus in most of the texts is predominantly 
conceptual. 

Summarizing, it seems apparent that cosmology is living a golden age with the advent of 
observations of high precision. Nonetheless, a critical revisiting of the direction in 
which it should go on appears also needed, for misconcepts like ``quantum backgrounds 
for classical cosmological settings'' and ``quantum gravity unification'' have not been 
decidedly achieved up-to-date. Thus, knowledge-building in cosmology, more than in any 
other field, should begin with visions of the reality, and passing to have a technical 
form whenever concepts and relations inbetween are translated into a mathematical 
structure. It is mandatory, therefore, that the meaning of such concepts be the same 
for all cosmologists, and that any relationship among all of them be tested both logically 
as well as mathematically. In other words, the notorius character of improbability
of  our universe, as is well-kown, assures to cosmologists a priviledged degree of 
freedom for formulating interpretations and theories. However, at the same time, it 
demands that their formulations and conclusions be considered in the light of data 
from astrophysical observations.

To the last, but not to end point, cosmology is living a moment in which open forums 
featured by authoritative scientists are inviting to admit that most current observations 
demand to revisit the pillars of the standard model. As a matter of examples, some years 
ago in ``The large-scale smoothness of the Universe'' by Wu, Lahav and Rees \cite{rees1999} 
have put into question the validity of the \textit{Cosmological Principle},
(which states that the universe is homogeneous and isotropic over cosmic distance scales),
under the argument that such pillar was put by hand by the founders of modern cosmology 
in an epoch where there were no conclusive astronomical observations that would provide 
the due support it had required. The reasons for revisiting the Cosmological Principle 
augmented over the last years after the verification of the existence of anomalies in 
the CMB maps obtained by WMAP, as pointed out by de Oliveira-Costa et al.\cite{angelica2003}
who verified that: a) the cosmic octupole on its own is anomalous at the 1-in-20 level by 
being very planar, and b) the alignment between the 4-pole and 8-pole is anomalous at the 
level 1-in-60 level. This result, which appears to be a fundamental one, has been 
addressed by other authors Schwarz et al.\cite{schwarz2004}, Land and Magueijo
\cite{land-magueijo2005,land-magueijo2007}, Wiaux et al.\cite{wiaux2006}, de 
Oliveira-Costa and Tegmark \cite{angelica-max2006}, Copi et al.\cite{copi2006}, 
Huterer\cite{huterer2006}, Cruz et al.\cite{cruz2006}, Bernui, Rebou\c cas and Tavakol
\cite{bernui2007}. Because of its relevance, these findings became matters of general
interest, as exemplified by a paper of scientific divulgation spirit: `` Why 
is the solar system cosmologically aligned?'' by Huterer \cite{huterer2007}, 
who signals out the ecliptic oddities ``The likelihood of these alignments happening 
by chance is less than 0.1 percent. Finally, the 4-pole and 8-pole planes are also 
perpendicular with the CMB dipole, which points to the direction of motion of the 
soslar system. Why CMB patterns are oriented to the solar system is not at all 
understood at this time''. Hence, bearing in mind that these results suggest a 
violation of the statistical isotropy (Cosmological Principle), one can conclude 
that such anomalies may  have a cosmological origin. If those results were to be 
explained in a conceptual and technical basis acceptable, and at the same time were 
excluded the possibility of having origin in a systematical error, then the relevance 
of Cosmological Principle for the standard model will have to be dismissed or at least 
partially limited.

Besides, a few years ago Fuzfa and Alimi \cite{alimi2006} showed that 
a self-consistent way to accommodate the SNIa observations goes through by admitting that 
the \textit{Equivalence Principle}, another pillar of the standard model, should be
forsaken. And even more, many years ago the Lema\^itre-Tolman-Bondi (LTB) cosmological
models were brought in into play as one can verify in the article by Hellaby \cite{hellaby2009} 
who reviewed those models which also give up at least the \textit{radial homogeneity} in 
cosmological scenarios. 
And to the last, several cosmological models recently idealized as an explanation of the 
alignments found in the observations by WMAP of the CMB radiation also invoke LTB models 
Caldwell and Stebbins \cite{cadwell2008}, or something of the like, and even a non scale 
invariant gravitational interaction \cite{cadwell2008A}, besides of calling also for 
cosmological models \textit{a l\'a} G\"odel. This moment of effervescence in cosmology 
invites to the reflexion and debate.

Perhaps we are already are in a position to successfully battle the "cosmic conundrum" 
and enter the golden age of cosmology. This is a moment to reflect on the advice given 
by Einstein~\cite{42}:

\begin{quotation}
``It has often been said, and certainly not without justification, that the man of science is a 
poor philosopher. Why then should it not be the right thing for the physicist to let the philosopher 
do the philosophizing? Such might indeed be the right thing at a time when the physicist believes he 
has at his disposal a rigid system of fundamental concepts and fundamental laws which are so well 
established that waves of doubt cannot reach them; but, it cannot be right at a time when the very 
foundations of physics itself have become problematic as they are now. At a time like the present, 
when experience forces us to seek a newer and more solid foundation, the physicist cannot simply 
surrender to the philosopher the critical contemplation of the theoretical foundations; for he 
himself knows best, and feels more surely where the shoe pinches. In looking for a new foundation, 
he must try to make clear in his own mind just how far the concepts which he uses are justified, 
and are necessities.''
\end{quotation}

\section*{Acknowledgments}

I am deeply thankful to Herman J. Mosquera Cuesta for his excellent 
comments, criticisms and suggestions that helped me to decidely improve the paper
up to reaching the present form. I am also graceful to Francisco Pelucio Silva for 
proofreading my English. And to both of them for their constant encouragement.


\begin{thebibliography}{99}
\bibitem{5}One could refer to this finding as the greatest discovery of all times.
\bibitem{6}It is common to find, in explanatory texts, the statement that, at this point, all 
galaxies, would be concentrated, which is not correct, even softening, the expression to the 
equivalent to all galaxies, simply because they did not exist yet.
\bibitem{13}This is the more used notation of the scalar factor $R(t)$, but many people use to write 
$a(t)$ instead. In this work, whenever the equation is presented in a historical order, we will 
preserve the notation used by the quoted author.
\end{thebibliography}
\end{document}